\documentclass[twocolumn,amssymb,nobibnotes,showpacs,superscriptaddress,aps,pra]{revtex4-1}
\usepackage{graphicx,amsmath,amssymb,epsfig,amsbsy}

\usepackage{epstopdf}

\begin{document}

\title{Manipulation and improvement of multiphoton blockade in a two cascade three-level atoms cavity-QED system}


\author{J. Z. Lin}
\affiliation{MOE Key Laboratory of Advanced Micro-Structured Materials,
	School of Physics Science and Engineering, Tongji University, Shanghai, China 200092}
\affiliation{Department of physics, Suzhou Institute of Technology, Jiangsu university of science and technology, Zhangjiagang, China 215600}
\author{K Hou}
\affiliation{MOE Key Laboratory of Advanced Micro-Structured Materials,
	School of Physics Science and Engineering,Tongji University, Shanghai, China 200092}
\affiliation{Department of Mathematics and Physics, Anhui JianZhu University, Hefei 230601,China}
\author{C. J. Zhu}
\email[Corresponding author:]{cjzhu@tongji.edu.cn}
\affiliation{MOE Key Laboratory of Advanced Micro-Structured Materials,
	School of Physics Science and Engineering,Tongji University, Shanghai, China 200092}
\author{Y. P. Yang}
\email[Corresponding author:]{yang\_yaping@tongji.edu.cn}
\affiliation{MOE Key Laboratory of Advanced Micro-Structured Materials,
	School of Physics Science and Engineering,Tongji University, Shanghai, China 200092}

\date{\today}

\begin{abstract}

We present a study of manipulating and improving the multiphoton blockade phenomenon in a single mode cavity with two cascade three-level atoms. Using an external control field, we show that the two-photon blockade can be changed to the three-photon blockade by just increasing the control field Rabi frequency when two atoms radiate in-phase. In the case of out-phase radiations, we show that the three-photon blockade can be significantly improved with an increased mean photon number when the control field is in the presence. The results presented in this work provide potential applications in quantum communication and quantum networking.
 
\end{abstract}

\pacs{42.50.Pq, 42.50.Nn, 37.30.+i}


\maketitle

\section{INTRODUCTION}
In an atom-cavity QED system driven by a coherent field, a single photon can suppress the transmission of other photons due to the strong coupling between the atom and cavity, which is well known as the two-photon blockade. This phenomenon is a close analogy to the phenomenon of Coulomb blockade ~\cite{Imamoglu}. After that, Birnbaum et al. demonstrate the photon blockade phenomenon experimentally in a single atom-cavity QED system with strong coupling strength, where the quantum statistic property of the incident photon stream was changed from the Poissonian distribution to sub-Poissonian distribution if the frequency of photons was tuned to one of the states of the lowest doublet dressed states~\cite{Birnbaum}. 

Due to it's potential applications in quantum communication and quantum networking, the study of two-photon blockade has received extensive attention in past few decades. A lot of experimental and theoretical works on photon blockade have been reported in various systems with strong coupling strength, including the circuit QED systems~\cite{Hoffman,Liu,Wang,Felicetti}, artificial atoms on a chip~\cite{Faraon,Reinhard}, optomechnical systems~\cite{Rabi,Ludwig,Nori,Hu,Xie,Wu}, and atom-cavity QED system~\cite{Hamsen,Schuster,Kubanek,Souza,Zhu,Werner,Rebic,Liao,Miranowicz}. Recently, 
the unconventional photon blockade based on quantum destructive in terference also gets a great deal of attention ~\cite{Liew} since the two-photon blockade effect can be significantly improved ~\cite{Majumdar,Bamba,Gerace,Flayac}. The corresponding experimental works were demonstrated in the quantum dot cavity QED system~\cite{Vaneph} and superconduncting circuit QED system~\cite{Snijders}, respectively.

Although the two-photon blockade has been studied extensively, the accomplishment of the multiphoton blocakde is challenging in experiments. A direct method to realize three-photon blockade is by increasing the incident field intensity so that two-photon excitations can be measured. However, the strong field intensity will result in the broadening of the dressed states, which compensates the inharmonic energy splitting and prohibits the observation of the three-photon blockade behavior. These characteristics have been observed by Hamsen et al.~\cite{Hamsen}. Another method to realize three-photon blockade is based on the collective decay of two atoms trapped in a single mode cavity with different coupling strengths~\cite{Zhu}. If two atoms have out-phase radiations, the three-photon blockade can be achieved since the two-photon excitations are dominant and one-photon excitations are forbidden. 

In this paper, we consider that two cascade three-level atoms are strongly coupled in a cavity, where two identical atoms interact with a pump field and a control field simultaneously. The motivation of adding a strong control field is to manipulate the dressed states of the system. We show that the photon blockade behavior can be changed and improved by adjusting the control field intensity. When two atoms radiate in-phase, the two-photon blockade can be changed to the three-photon blockade by in creaing the control field Rabi frequency. In the case of out-phate, we show that the three-photon blockade can be remarkably improved in the presence of the control field. 

\section{model and dressed state picture}
\begin{figure}[htbp]
	\centering
	\includegraphics[width=8.5cm]{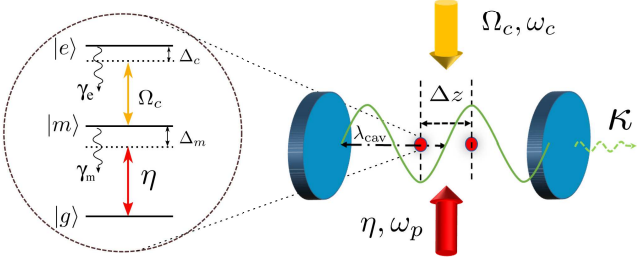}
	\caption{(Color online) Schematic of the two atoms cavity QED system.  The wavelength of this single model cavity is $\lambda_{\rm cav}$, and the corresponding angular frequency is $\omega_{\rm cav}=2\pi C/\lambda_{\rm cav}$. Each atom has three energy levels labeled as $|g\rangle$, $|m\rangle$ and $|e\rangle$, respectively. A coherent pump field with Rabi frequency $\eta$ (angular frequency is $\omega_p$) drives the $|g\rangle\leftrightarrow|m\rangle$ transition, and a strong control field with Rabi frequency $\Omega_c$ (angular frequency is $\omega_c$) drives the $|m\rangle\leftrightarrow|e\rangle$ transition. The one-photon detunings are defined as $\Delta_m=\omega_m-\omega_g-\omega_p$ and $\Delta_c=\omega_e-\omega_m-\omega_c$. Here, the distance between two atoms is $\Delta z$. The spontaneous emission rate of state $|m\rangle$ ($|e\rangle$) is $\gamma_m$ ($\gamma_e$), and the cavity decay rate is $\kappa$.}
	\label{Fig.1}
\end{figure}
As depicted in Fig.~\ref{Fig.1}, this two-atom cavity QED system consists of a single-mode cavity and two identical cascade three-level atoms directly driven by a pump field $\eta$ and a control field $\Omega_c$ simultaneously. We assume that the pump (control) field drives the  $|g\rangle\leftrightarrow|m\rangle$ ($|m\rangle\leftrightarrow|e\rangle$) transition, while the cavity mode only couples the $|g\rangle\leftrightarrow|m\rangle$. Clearly, the system is the same as that proposed in Ref.~\cite{Zhu} when the control field is turned off (i.e., setting $\Omega_c=0$).  

In general, the dynamical behavior of this cavity QED system can be described by using the master equation, i.e.,
\begin{equation}\label{1}
	\frac{d\rho}{dt}=-\frac{i}{\hbar}[H,\rho]+\mathcal{L}_\kappa\rho+\mathcal{L}_\gamma\rho,
\end{equation}
where $\rho$ is the density-matrix operator of the atom-cavity QED system. Under the rotating-wave and electric dipole approximations, the system Hamiltonian can be written as $H=H_0+H_I+H_L$ with
\begin{align*}
	&H_0=\hbar\sum_{i=1,2}(\Delta_e\sigma_{ee}^i+\Delta_m\sigma_{mm}^i+\Delta_{\rm cav} a^\dag a),\\
	&H_I=\hbar\sum_{i=1,2}g_i(a\sigma_{mg}^i+a^\dag\sigma_{gm}^i),\\
	&H_L=\hbar[\eta\sum_{i=1,2}(\sigma_{mg}^i+\sigma_{gm}^i)+\Omega_c\sum_{i=1,2}(\sigma_{me}^i+\sigma_{em}^i)],
\end{align*}
where $H_0$ is the energy of atoms and the cavity field, $H_I$ represents the interaction between atoms and the cavity field, and $H_L$ is the coherent driving term involving the pump field and control field. The coupling strength between the $i$-th atom and cavity $g_i=g\cos(2\pi z_i/\lambda_{\rm cav})$ is dependent to the position of the $i$-th atom $z_i$, where $\lambda_{\rm cav}$ is the wavelength of the cavity mode (the corresponding angular frequency is $\omega_{\rm cav}=2\pi c/\lambda_{\rm cav}$). Here, $a$ and $a^\dag$ are the annihilation and creation operators of the cavity mode, respectively. $\sigma_{jk}^i=|j\rangle^i\langle k|$ ($j,k=\{g,m,e\}$) denotes the atomic operator of the $i$-th atom. The detunings are defined as $\Delta_{\rm cav}=\omega_{\rm cav}-\omega_p$, $\Delta_m=\omega_m-\omega_g-\omega_p$, $\Delta_e=\omega_e-\omega_g-(\omega_p+\omega_c)=\Delta_m+\Delta_c$ with $\Delta_c=\omega_e-\omega_m-\omega_c$ and $\hbar\omega_j$ ($j=\{g,m,e\}$) being the energy of state $|j\rangle$. 

The last two terms in Eq.~(\ref{1}) denote the decay of the atom and cavity, which are given by
\begin{align}
	\mathcal{L}_{\rm cav}\rho=&\kappa(2a\rho a^\dag-a^\dag a\rho-\rho a^\dag a),\\
	\mathcal{L}_{\rm atom}\rho=&\sum_{i=1,2}\left[\gamma_e(2\sigma_{me}^i\rho\sigma_{em}^{i}-\sigma_{em}^{i}\sigma_{me}^i\rho-2\rho\sigma_{em}^{i}\sigma_{me}^i)\right.\nonumber\\
	&\left.+\gamma_m(2\sigma_{gm}^i\rho\sigma_{mg}^{i}-\sigma_{mg}^{i}\sigma_{gm}^i\rho-2\rho\sigma_{mg}^{i}\sigma_{gm}^i),\right],
\end{align}
respectively. Here $\kappa$ is the cavity decay rate, and $\gamma_{\alpha}$ ($\alpha=m,e$) is the spontaneous emission rate of the state $|\alpha\rangle$.

To understand the physical mechanism clearly, we rewrite the Hamiltonian of the system in dressed state picture by using $|GG,n\rangle$, $|MG\pm,n-1\rangle$, $|EG\pm,n-1\rangle$, $|MM,n-2\rangle$, $|EM\pm, n-2\rangle$ and $|EE,n-2\rangle$ as basis in $n-$photon space (the definition of these basis are given in the appendix).  Assuming $\Delta_c=0$ and $\omega_{\rm cav}=\omega_m-\omega_g$ for mathematical simplicity, we have $\Delta_m=\Delta_e=\Delta_{\rm cav}\equiv\Delta_p$. Under the weak pump field approximation, the effects of the pump field can be treated as a perturbation to the system. Then, the Hamiltonian in one-photon space is expressed as
\begin{eqnarray}\label{eq:H1ph}
	H_{\rm 1ph}=\left(
	\begin{array}{ccccc}
		0 & g_+/\sqrt{2} & g_-/\sqrt{2} & 0 & 0 \\
		g_+/\sqrt{2} & 0 & 0 & \Omega_c  & 0 \\
		g_-/\sqrt{2} & 0 & 0 & 0 & \Omega_c  \\
		0 & \Omega_c  & 0 & 0 & 0 \\
		0 & 0 & \Omega_c  & 0 & 0 \\
	\end{array}
	\right),
\end{eqnarray}
and, in two-photon space, the matrix of the Hamiltonian is given by
\begin{eqnarray}\label{eq:H2ph}
&&H_{\rm 2ph}=\nonumber\\
&&\left(
\begin{array}{ccccccccc}
0 & g_+ & g_- & 0 & 0 & 0 & 0 & 0 & 0 \\
g_+ & 0 & 0 & \Omega_c  & 0 & 0 & 0 & \frac{g_+}{\sqrt{2}} & 0 \\
g_- & 0 & 0 & 0 & \Omega_c  & 0 & 0 & -\frac{g_-}{\sqrt{2}} & 0 \\
0 & \Omega_c  & 0 & 0 & 0 & \frac{g_+}{2} & -\frac{g_-}{2} & 0 & 0 \\
0 & 0 & \Omega_c  & 0 & 0 & -\frac{g_-}{2} &\frac{g_+}{2} & 0 & 0 \\
0 & 0 & 0 & \frac{g_+}{2} & -\frac{g_-}{2} & 0 & 0 & \sqrt{2} \Omega_c  & \sqrt{2} \Omega_c  \\
0 & 0 & 0 & -\frac{g_-}{2} & \frac{g_+}{2} & 0 & 0 & 0 & 0 \\
0 & \frac{g_+}{\sqrt{2}} & -\frac{g_-}{\sqrt{2}} & 0 & 0 & \sqrt{2} \Omega_c  & 0 & 0 & 0 \\
0 & 0 & 0 & 0 & 0 & \sqrt{2}\Omega_c  & 0 & 0 & 0 
\end{array}
\right),\nonumber\\
\end{eqnarray}

where $g_\pm=g_1\pm g_2=g(1\pm\cos{\phi_z})$ with $\phi_z=2\pi\Delta z/\lambda_{\rm cav}$.

\begin{figure}[htbp]
	\centering
	\includegraphics[width=8.5cm]{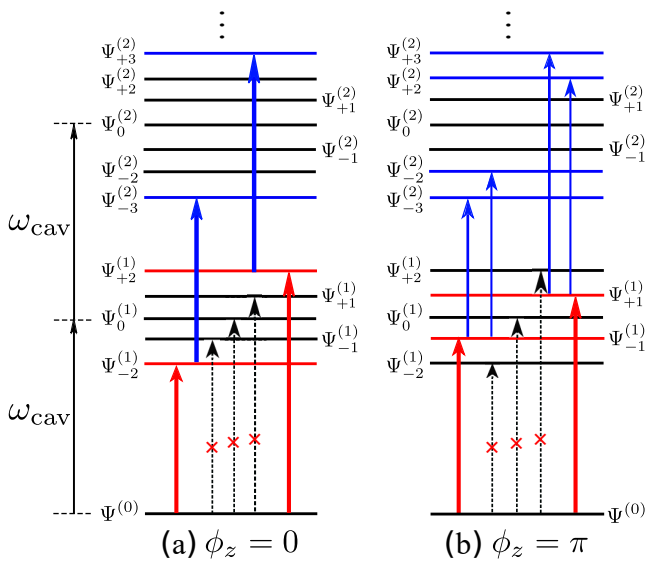}
	\caption{(Color online) The dressed-state energy structures for $\phi_z=0$ [panel (a)] and $\phi_z=\pi$ [panel (b)], respectively. The red arrows represent the one-photon transitions, but the blue ones represent the two-photon transitions. Here, we only show several main pathways of two photon transitions. The black arrows with red cross denote that the transitions are forbidden.}
	\label{Fig.2}
\end{figure}

Diagonalizing Eqs.~(\ref{eq:H1ph}) and (\ref{eq:H2ph}), we can obtain the corresponding eigenvalues and eigenstates forming the dressed states in one- and two-photon space (see Appendix). Furthermore, we can also obtain the transition strength by calculating the operator $\eta\sum_{i=1,2}(\sigma_{mg}^i+\sigma_{gm}^i)$. These dressed states along with some of the important transitions are shown in Fig.~\ref{Fig.2} and the physical mechanism of the photon blockade can be understand easily. For example, in the case of $\phi_z=0$ [i.e., panel(a)], the system can only absorb a single photon to be excited to the one-photon states such as $\Psi^{(1)}_{\pm2}$. However, the second photon can`t be absorbed due to the inharmonic energy splitings. This phenomenon is widely known as the two-photon blockade. In the case of $\phi_z=\pi$ [i.e., panel(b)], the states in the one-photon space (i.e., $n=1$) aren`t allowed to be excited because of the destructive interference ~\cite{Zhu}. Thus, the system can only be excited to the states in the two-photon space (i.e., $n=2$) via two photon processes (e.g., $\Psi^{(0)}\to \Psi^{(1)}_{-1}\to\Psi^{(2)}_{-2(-3)}$). Likewise, the third photon can`t be absorbed by the system due to the energy difference. This phenomenon is rrecognized as the three-photon blockade ~\cite{Zhu}. In the following, we will show how the control field affects the photon blockade phenomena. 

\section{THE MANIPULATION OF PHOTON BLOCKADE}
Now, we focus on the case that the two atoms have the same coupling strengths, i.e., $g_1=g_2=g$ leading to $\phi_z=0$. To show the diffeences between this three-level atom cavity-QED system and the typical two-level atom cavity-QED system, we Numerically solving  Eq.~(\ref{1}) under the condition of a weak pump field. 
\begin{figure}[htbp]
	\centering
	\includegraphics[width=8.5cm]{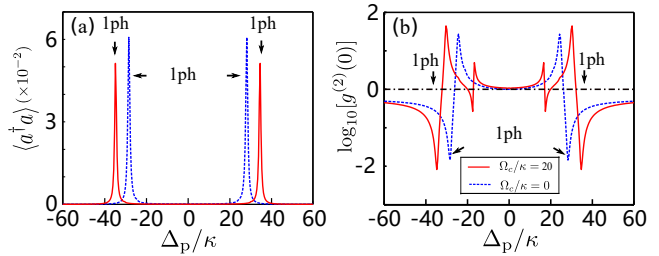}
	\caption{(Color online) Panels(a) and (b) display the mean photon number $\langle a^\dag a\rangle$ and the equal-time photon-photon correlation function $\log_{10}[g^{(2)}(0)]$ as a function of the normalized detuning $\Delta_p/\kappa$, respectively. The Rabi frequency of the control field is chosen as $\Omega_c/\kappa=0$ (blue dashed curves) and $20$ (red solid curves), respectively. Other system parameters are given by $\eta/\kappa=0.2$, $g/\kappa=20$, $\gamma_m/\kappa=1$ and $\gamma_e/\kappa=0.01$. The black dash-dotted line in panel (b) indicates $\log_{10}[g^{(2)}(0)]=0$.}
	\label{Fig.3}
\end{figure}

As shown in Fig.~\ref{Fig.3}, the mean photon number $\langle a^\dag a\rangle$ [panel (a)] and the equal-time photon-photon correlation function $g^{(2)}(0)=\langle a^\dag a^\dag aa\rangle/\langle a^\dag a\rangle^2$ in logarithmic unit [panel (b)] are ploted as a function of the normalized detuning $\Delta_p/\kappa$. Here we choose the control field Rabi frequency as $\Omega_c/\kappa=0$ (i.e., two-level system, blue dashed curves) and $\Omega_c/\kappa=20$ (red solid curves), respectively. Other system parameters are chosen as $\eta/\kappa=0.2$, $g/\kappa=20$,$\gamma_m/\kappa=1$ and $\gamma_e/\kappa=0.01$. For $\Omega_c=0$, one can observe two peaks in the cavity excitation spectrum at $\Delta_p=\pm\sqrt{2}g$, corresponding to the frequencies of one-photon excitations (i.e., $\Psi^{(0)}\rightarrow\Psi_{\pm2}^{(1)}$ transitions, see the blued dashed curve in panel (a)). Since the pump field is very weak and one-photon excitations are dominant, the quantum features of the cavity feild can be characterized by the second-order correlation function $g^{(2)}(0)$. As shown in Fig.~\ref{Fig.3}(b), the value of $g^{(2)}(0)$ at one-photon excitation is smaller than unity (i.e., $\log_{10}[g^{(2)}(0)]<0$), which implies that the two-photon blockade behavior and the nonclassical cavity field with sub-Poissonian distribution can be achieved.

For $\Omega_c\ne0$, the energies of the states $\Psi_{\pm2}^{(1)}$ shift as the control field intensity increases (see Fig.~2(a) and formulas in the appendix). As a result, the width between two peaks in the cavity excitation spectrum$\Gamma_{\rm w}=2\sqrt{2g^2+\Omega_c^2}$ becomes larger than that in the absence of the control field (see the red solid curve in panel (a)). Compared with the case of $\Omega_c=0$, the minimum value of $g^{(2)}(0)$ will decrease, corresponding to a slight improvement of the two-photon blockade. The physical mechanism of this improvement attributes to two factors: the energy shifting of the dressed states and the transition strengths. The former causeds an increase of the energy difference between states $\Psi_{\pm2}^{(1)}$ and $\Psi_{\pm3}^{(2)}$ (the explicit expression is given in the appendix). Therefore, the second photon becomes much harder to be absorbed by the system. Obviously, adjusting the control field Rabi frequency, one can control the frequency for realizing the two-photon blockade. 

Next, we consider the strong pumping case (e.g.,$\eta/\kappa=1.5$), where the two-photon excitations are strong as the one-photon excitations. 
\begin{figure}[htbp]
	\centering
	\includegraphics[width=8.5cm]{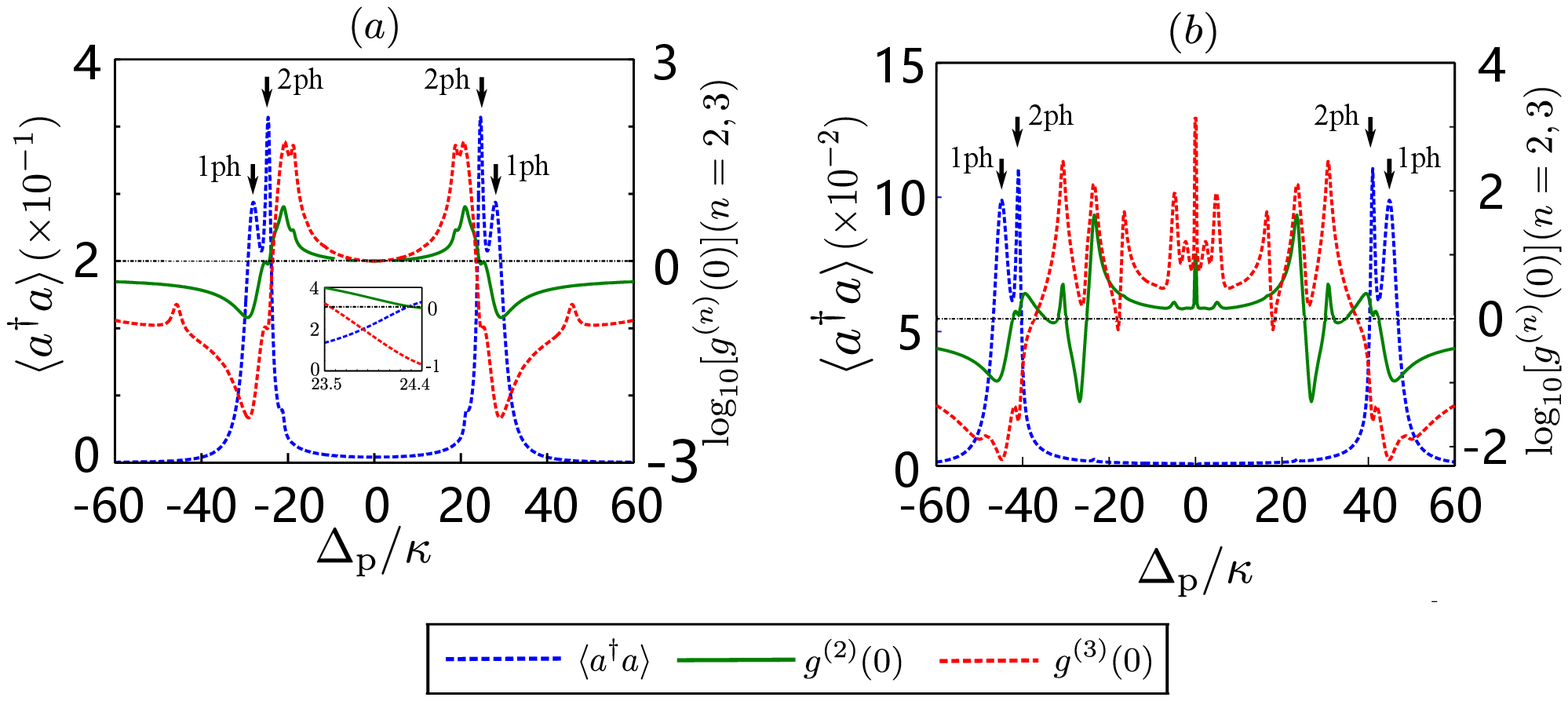}
	\caption{(Color online) Plots of the mean photon number (blue dashed curves), equal-time second-order field correlation function $g^{(2)}(0)$ (green solid curves) and third-order field correlation function $g^{(3)}(0)$ (red dashed curves). In panels (a) and (b), the control field Rabi frequency is chosen as $\Omega_c/\kappa=0$ and $\Omega_c/\kappa=35$, respectively. The inserted plot in panel (a) shows the regime of two photon excitation, and the black dash-dotted lines indicate $g^{(2)}(0)=g^{(3)}(0)=1$. Here, we choose $\eta/\kappa=1.5$, $\phi_z=0$ and other system parameters are the same as those used in Fig.~\ref{Fig.3}.}
	\label{Fig.4}
\end{figure}
In the absence of the control field, there are four peaks in the cavity excitation spectrum (see Fig.~\ref{Fig.4}(a)), corresponding to the frequencies $\Delta_p=\pm\sqrt{2}g$ (one-photon excitations) and $\Delta_p=\pm\sqrt{6}g/2$ (two-photon excitations), respectively~\cite{Zhu}. At frequencies of one-photon excitations, one can obtain $g^{(2)}(0)\approx0.2$, i.e., the two-photon blockade. Here, the system parameters are the same as those used in Fig.~\ref{Fig.3}. Near the two-photon excitation frequencies, it is possible to observe the three-photon blockade (i.e., $g^{(2)}(0)>1$ and $g^{(3)}(0)<1$) in a narrow regime (see the inserted plot in panel(a)). It is noted that the observation of the three-photon blocakde is a challenging in experiments and technique noise becomes a fatal problem because the frequency of the pump field must bu controlled precisely. In the presence of  the control field ($\Omega_c/\kappa=35$), the frequencies for realizing the two- and three-photon blockade phenomena are changed because of the energy shiftings of the dressed states [see Fig.~\ref{Fig.4}(b) and the appendix]. 
the energy shifts cause a significant improvement of the photon blockade phenomenon (see panel (b)). At the frequencies of one-photon excitation, it is found that $g^{(2)}(0)\approx0.136$, which is near $2$ times smaller than that in the absence of the control field. Moreover, at the frequencies of two-photon excitations, the frequency regime for realizing the three-photon blockade regime is much broader than the case of $\Omega_c=0$. For example, one can obtain $g^{(2)}(0)\approx1.13$ and $g^{(3)}(0)\approx0.025$ with mean photon number $n=0.11$ at the frequency of two-photon excitation, which implies the three-photon blockade behavior.

\section{SIGNIFICANT IMPROVEMENT OF THE THREE-PHOTON BLOCKADE}
\begin{figure}[htbp]
	\centering
	\includegraphics[width=8.5cm]{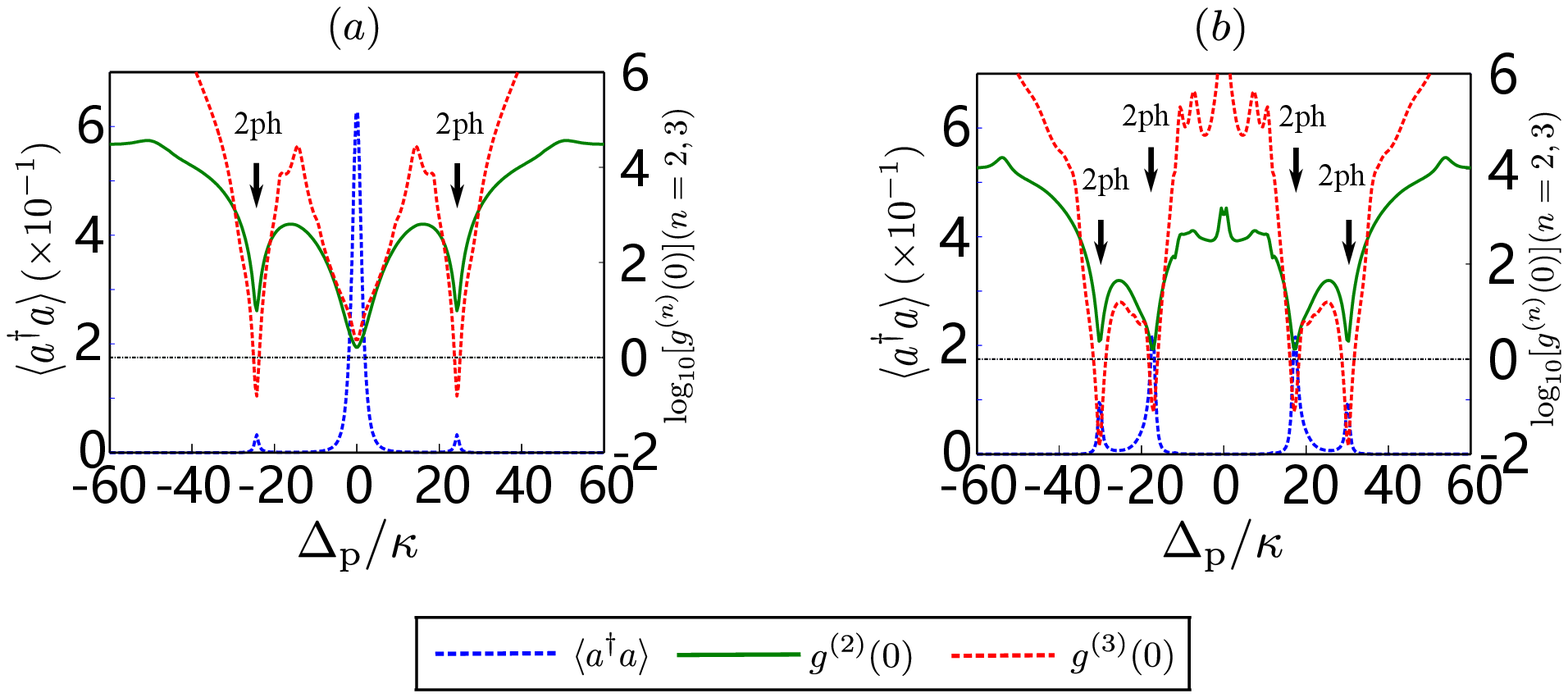}
	\caption{(Color online) Panels (a) and (b): Plots of the mean photon number (blue dashed curves), equal-time second-order (green solid curves) and third-order (red dashed curve) field correlation functions for $\Omega_c/\kappa=0$ (a) and $\Omega_c/\kappa=20$ (b). Here, we choose $\eta/\kappa=2$, $\phi_z=\pi$ and other system parameters are the same as those used in Fig. 3. The black dash-dotted line indicates $g^{(2)}(0)=g^{(3)}(0)=1$.}
	\label{Fig.5}
\end{figure}

Finally, we consider that two atoms have different coupling strengths (i.e., $g_1=-g_2=g$ leading to $\phi_z=\pi$). In this case,  the two-photon excitation becomes dominant because $\Psi_0\to\Psi^{(1)}_{\pm2}$ transitions are not allowed ~\cite{Zhu}. As a result, two side peaks (see Fig.~\ref{Fig.5}(a)), corresponding to the two-photon excitations $\Psi_0\to\Psi^{(1)}_{\pm1}\to\Psi^{(2)}_{\pm3}$ with $\Delta_p=\pm\sqrt{6}g/2$, can be observed in the cavity excitation spectrum when the control field is absent (i.e., $\Omega_c=0$).
Correspondingly, the three-photon blockade  ($g^{(2)}(0)\approx9.48>1$ and $g^{(3)}(0)\approx0.14<1$) occurs at the two-photon excitation frequencies. We also note that the central peak in the cavity excitation spectrum arises from the multiphoton excitation process~\cite{Zhu}, which results in a classical field generation (i.e., $g^{(2)}(0)>1$ and $g^{(3)}(0)>1$). In the presence of the control field(e.g., $\Omega_c/\kappa=20$), there exist four peaks in the cavity excitation spectrum as shown in Fig.~\ref{Fig.5}(b). Examining the dressed states [see Fig. 2(b) and the appendix], all of them attribute to the two-photon excitations: $\Psi_0\to\Psi^{(1)}_{\pm1}\to\Psi^{(2)}_{\pm2}$ with $\Delta_p/\kappa=\pm\sqrt{\alpha-\sqrt{\beta}}/(2\sqrt{2})$ and $\Psi_0\to\Psi^{(1)}_{\pm1}\to\Psi^{(2)}_{\pm3}$ with $\Delta_p/\kappa=\pm\sqrt{\alpha+\sqrt{\beta}}/(2\sqrt{2})$. Here, the variables $\alpha$ and $\beta$ are functions of the control field Rabi frequency $\Omega_c$ (see appendix). Compared with the case of $\Omega_c/\kappa=0$, it is clear to see that the mean photon number is enhanced, and the three-photon blockade phenomenon can be significantly improved. At the detuning $\Delta_p/\kappa=\pm\sqrt{\alpha+\sqrt{\beta}}/(2\sqrt{2})$, one can obtain $g^{(2)}(0)\approx2.37$ and $g^{(3)}(0)\approx0.015$ with mean photon number $n=0.1$. We must point out that the value of $g^{(3)}(0)$ is about 10 times smaller than that in the case of $\Omega_c/\kappa=0$. The physical mechanism can also be explained by examining the energy difference between the dressed states, where the states in the three-photon space must be considered because of the three photon processes. In this case, the energy differences for the three-photon absorption $\Delta E^{\pm}_{3ph}$ and $\Delta E^{`\pm}_{3ph}$ gradually increases as the control field intensity increases [see Fig.6 in the appendix], which prohitbits the absorption of the third photons and leads to a significant improvement of the three-photon blockade.

\section{Conclusion}
In summary, we have studied the quantum properties of the cavity field in the atom-cavity QED system with two cascade three-level atoms driven by a pump field and a control field simultaneously. we show that the quantum fluctuation of the cavity field can be controlled by tuning the control field intensity since the dressed states are shifted by the control field. When two atoms are in-phase radiations, for example, we show that the frequency to realize the two-photon blockade can be actively controlled by adjusting the control field intensity since the dressed states are shifted by the control field. We also show that, increasing the pump field Rabi frequency, not only two-photon but also three-photon blockades can be observed. In the case of out-phase radiations, we show that the three-photon blockade phenomenon can be significantly improved with enhanced photon numbers. These may result in possible applications in quantum communication and quantum networking.
   
\begin{acknowledgments}
We acknowledge the National Key Basic Research Special Foundation (Grant No. 2016YFA0302800); the Shanghai Science and Technology Committee (Grant No. 18JC1410900); the National Nature Science Foundation (Grant No. 11774262); the Anhui Provincial Natural Science Foundation (Grant No.1608085QA23) 
\end{acknowledgments}

\appendix


\section{The eigenvalues and eigenstates of Eqs. (4) and (5)}
The Hamiltonian of the system in dressed state picture can be rewritten by using $|GG,n\rangle$, $|MG\pm,n-1\rangle$, $|EG\pm,n-1\rangle$, $|MM,n-2\rangle$, $|EM\pm, n-2\rangle$ and $|EE,n-2\rangle$ as basis in $n-$photon space, which are defined as
\begin{align*}
&|GG,n\rangle=|gg,n\rangle\\
&|MG\pm,n-1\rangle=\frac{1}{\sqrt{2}}(|mg,n-1\rangle\pm|gm,n-1\rangle)\\
&|EG\pm,n-1\rangle=\frac{1}{\sqrt{2}}(|eg,n-1\rangle\pm|ge,n-1\rangle)\\
&|MM,n-2\rangle=|mm,n-2\rangle\\
&|EM\pm,n-2\rangle=\frac{1}{\sqrt{2}}(|em,n-2\rangle\pm|me,n-2\rangle)\\
&|EE,n-2\rangle=|ee,n-2\rangle.
\end{align*}

\vskip 5pt
\noindent{\it Case 1: $\phi_z=0$.} Diagonalizing the Eq.~(4), we  can obtain the eigenvalues of the dressed states in one-photon space, i.e., $\lambda^{(1)}_0=0$, $\lambda_{1\pm}^{(1)}=\pm\Omega_c$ and $\lambda_{2\pm}^{(1)}=\pm\sqrt{2g^2+\Omega_c^2}$, respectively. The corresponding eigenstates are given by
\begin{align*}   
&\Psi_0^{(1)}=\frac{-\Omega_c}{\sqrt{2}g}|GG,1\rangle+|EG+,0\rangle\\
&\Psi_{1\pm}^{(1)}=\pm|MG-,0\rangle+|EG-,0\rangle\\
&\Psi_{2\pm}^{(1)}=\frac{\sqrt{2}g}{\Omega_c} |GG,1\rangle\pm\frac{\sqrt{2g^2+\Omega_c^2}}{\Omega_c}|MG+,0\rangle+|EG+,0\rangle
\end{align*}

In the two-photon space, the eigenvalues can be obtained by solving Eq.~(5), yielding $\lambda_{0}^{(2)}=\lambda_{0\pm}^{(2)}=0$, $\lambda_{1\pm}^{(2)}=\pm\sqrt{g^2+\Omega_c^2}$, $\lambda_{2\pm}^{(2)}=\pm\sqrt{\alpha-\sqrt{\beta}}/\sqrt{2}$ and $
\lambda_{3\pm}^{(2)}=\pm\sqrt{\alpha+\sqrt{\beta}}/\sqrt{2}$ with $\alpha=7g^2+5\Omega_c^2$ and $\beta=25g^4+6g^2\Omega_c^2+9\Omega_c^4$. Correspondingly, the eigenstates are given by
\begin{align*}
\Psi_{0+}^{(2)}=&\frac{\Omega_c^2}{\sqrt{2}g^2}|GG,2\rangle-\frac{\sqrt{2}\Omega_c}{g}|EG+,1\rangle+|EE,0\rangle\\
\Psi_{0-}^{(2)}=&\frac{\Omega_c^2-g^2}{\sqrt{2}g^2}|GG,2\rangle-\frac{\sqrt{2}\Omega_c}{g}|EG+,1\rangle+|MM,0\rangle\\
\Psi_{0}^{(2)}=&\frac{-g}{\Omega_c}|MG-,1\rangle+|EM-,0\rangle\\
\Psi_{1\pm}^{(2)}=&\frac{\Omega_c}{g}|MG-,1\rangle\pm\frac{\sqrt{g^2+\Omega_c^2}}{g}|EG-,1\rangle+|EM-,o\rangle\\   
\Psi_{2\pm}^{(2)}=&\pm\frac{\gamma+\sqrt{\beta}}{3\sqrt{2}\Omega_c^2}|GG,2\rangle\mp\frac{\sqrt{\alpha-\sqrt{\beta}}(\gamma+\sqrt{\beta})}{12g\Omega_c^2}|MG+,1\rangle\nonumber\\
&-\frac{(\gamma-6g^2)+\sqrt{\beta}}{6\sqrt{2}g\Omega_c}|EG+,1\rangle\pm\frac{\sqrt{\alpha-\sqrt{\beta}}}{2\Omega_c}|EM+,0\rangle\nonumber\\
&+\frac{-\gamma+6\Omega_c^2-\sqrt{\beta}}{6\Omega_c^2}|MM,0\rangle+|EE,0\rangle\\   
\Psi_{3\pm}^{(2)}=&\frac{-\gamma+\sqrt{\beta}}{3\sqrt{2}\Omega_c^2}|GG,2\rangle  \pm\frac{\sqrt{\alpha+\sqrt{\beta}}(-\gamma+\sqrt{\beta})}{12g\Omega_c^2}|MG+,1\rangle\nonumber\\
&+\frac{(-\gamma+6g^2)+\sqrt{\beta}}{6\sqrt{2}g\Omega_c}|EG+,1\rangle   \pm\frac{\sqrt{\alpha+\sqrt{\beta}}}{2\Omega_c}|EM+,0\rangle\nonumber\\
&+\frac{-\gamma+6\Omega_c^2+\sqrt{\beta}}{6\Omega_c^2}|MM,0\rangle  +|EE,0\rangle,
\end{align*}
with $\gamma=-5g^2+3\Omega_c^2$. 

\begin{figure}[h!]
	\centering
	\includegraphics[width=\linewidth]{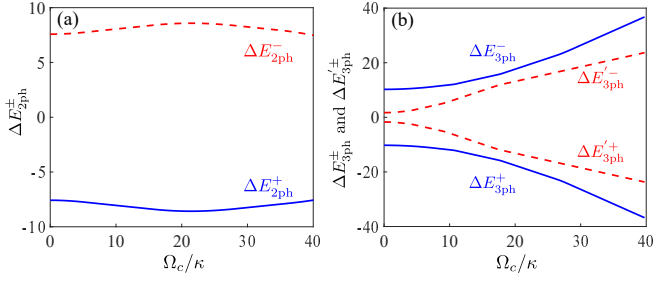}
	\caption{(Color online) Then energy differences for two-photon process [panel (a)] and three-photon process [panel (b)]. Here, the coupling strength is $g/\kappa=20$ and other system parameters are the same as those used in Fig. 3. }
	\label{Fig.S1}
\end{figure}
We note that the second pump field photon can't be absorbed at the detuning $\Delta_p=\lambda^{(1)}_{2\pm}$ due to the energy difference between the states $\Psi_{2\pm}^{(1)}$ and $\Psi_{3\pm}^{(2)}$, which is given by $\Delta E^{\pm}_{\rm 2ph}=\lambda_{3\pm}^{(2)}-2\lambda_{2\pm}^{(1)}$. As shown in Fig.~\ref{Fig.S1}(a), the energy differences increase slightly at the beginning, but drop quickly as the control field Rabi frequency increases. Thus, the improvement of the two-photon blockade is very weak as shown in Fig. 4(a) and 4(b).

\vskip 5pt
\noindent{\it Case 2: $\phi_z=\pi$.}, In the one-photon space, the eigenvalues are given by $\lambda_0^{(1)}=0$, $\lambda_{1\pm}^{(1)}=\pm\Omega_c$ and $\lambda_{2\pm}^{(1)}=\pm\sqrt{2g^2+\Omega_c^2}$ with the corresponding eigenstates
\begin{align*}
&\Psi_{0}^{(1)}=-\frac{\Omega_c}{\sqrt{2}g}|GG,1\rangle+|EG-,0\rangle\\
&\Psi_{1\pm}^{(1)}=\pm|MG+,0\rangle+|EG+,0\rangle\\
&\Psi_{2\pm}^{(1)}=\frac{\sqrt{2}g}{\Omega_c}|GG,1\rangle\pm\frac{\sqrt{2g^2+\Omega_c^2}}{\Omega_c}|MG-,0\rangle+|EG-,0\rangle
\end{align*}

In the two-photon space, we can obtain $\lambda_{0}^{(2)}=\lambda_{0\pm}^{(2)}=0$, $\lambda_{1\pm}^{(2)}=\pm\sqrt{g^2+\Omega_c^2}$, $\lambda_{2\pm}^{(2)}=\pm\frac{\sqrt{\alpha-\sqrt{\beta}}}{\sqrt{2}}$ and $\lambda_{3\pm}^{(2)}=\pm\frac{\sqrt{\alpha+\sqrt{\beta}}}{\sqrt{2}}$. The corresponding eigenstates are given by
\begin{align*}
\Psi_{0+}^{(2)}=&-\frac{\Omega_c^2}{\sqrt{2}g}|GG,2\rangle+\frac{\sqrt{2}\Omega_c}{g}|EG-,1\rangle+|EE,0\rangle\\
\Psi_{0-}^{(2)}=&\frac{g^2-\Omega_c^2}{\sqrt{2}g}|GG,2\rangle+\frac{\sqrt{2}\Omega_c}{g}|EG-,1\rangle+|MM,0\rangle\\
\Psi_{0}^{(2)}=&\frac{g}{\Omega_c}|MG+,1\rangle+|EM-,0\rangle\\
\Psi_{1\pm}^{(2)}=&-\frac{\Omega_c}{g}|MG+,1\rangle\mp\frac{g^2+\Omega_c^2}{g}|EG+,1\rangle+|EM-,0\rangle\\
\Psi_{2\pm}^{(2)}=&\frac{\gamma+\sqrt{\alpha}}{3\sqrt{2}\Omega_c^2}|GG,2\rangle\pm\frac{\sqrt{\alpha-\sqrt{\beta}}(\gamma+\sqrt{\beta})}{12g\Omega_c^2}|MG-,1\rangle\nonumber\\
&+\frac{\gamma-6g^2+\sqrt{\beta}}{6\sqrt{2}g\Omega_c}|EG-,1\rangle\pm\frac{\sqrt{\alpha-\sqrt{\beta}}}{2\Omega_c}|EM+,0\rangle\nonumber\\
&+\frac{-\gamma+6\Omega_c^2-\sqrt{\beta}}{6\Omega_c^2}|MM,0\rangle+|EE,0\rangle\\ 
\Psi_{3\pm}^{(2)}=&-\frac{-\gamma+\sqrt{\beta}}{3\sqrt{2}\Omega_c^2}|GG,2\rangle\mp\frac{\sqrt{\alpha+\sqrt{\beta}}(-\gamma+\sqrt{\beta})}{12g\Omega_c^2}|MG-,1\rangle\nonumber\\
&-\frac{-\gamma+6g^2+\sqrt{\beta}}{6\sqrt{2}g\Omega_c}|EG-,1\rangle\pm\frac{\sqrt{\alpha+\sqrt{\beta}}}{2\Omega_c}|EM+,0\rangle\nonumber\\
&+\frac{-\gamma+6\Omega_c^2+\sqrt{\beta}}{6\Omega_c^2}|MM,0\rangle+|EE,0\rangle
\end{align*}

To discuss the three-photon blockade, the dressed states in the three-photon space must be considered. In the three-photon space, the Hamiltonian is written as
\begin{widetext}
\begin{eqnarray*}\label{eq:H3ph}
&&H_{\rm 3ph}=\left(
\begin{array}{ccccccccc}
0 & \frac{\sqrt{6}g_+}{2} &  \frac{\sqrt{6}g_-}{2} & 0 & 0 & 0 & 0 & 0 & 0 \\
\frac{\sqrt{6}g_+}{2} & 0 & 0 & \Omega_c  & 0 & 0 & 0 & g_+ & 0 \\
\frac{\sqrt{6}g_-}{2} & 0 & 0 & 0 & \Omega_c  & 0 & 0 & -g_- & 0 \\
0 & \Omega_c  & 0 & 0 & 0 & \frac{\sqrt{2}g_+}{2} & -\frac{\sqrt{2}g_-}{2} & 0 & 0 \\
0 & 0 & \Omega_c  & 0 & 0 & -\frac{\sqrt{2}g_-}{2} &\frac{\sqrt{2}g_+}{2} & 0 & 0 \\
0 & 0 & 0 & \frac{\sqrt{2}g_+}{2} & -\frac{\sqrt{2}g_-}{2} & 0 & 0 & \sqrt{2} \Omega_c  & \sqrt{2} \Omega_c  \\
0 & 0 & 0 & -\frac{\sqrt{2}g_-}{2} & \frac{\sqrt{2}g_+}{2} & 0 & 0 & 0 & 0 \\
0 & g_+ & -g_- & 0 & 0 & \sqrt{2} \Omega_c  & 0 & 0 & 0 \\
0 & 0 & 0 & 0 & 0 & \sqrt{2}\Omega_c  & 0 & 0 & 0 
\end{array}
\right).\nonumber\\
\end{eqnarray*}
\end{widetext}
The eigenvalues of the above matrix are  $\lambda_{0}^{(3)}=\lambda_{0\pm}^{(3)}=0$, $\lambda_{1\pm}^{(3)}=\pm\sqrt{2g^2+\Omega_c^2}$, $\lambda_{2\pm}^{(3)}=\pm\sqrt{X-\sqrt{Y}}/\sqrt{2}$ and $
\lambda_{3\pm}^{(3)}=\pm\sqrt{X+\sqrt{Y}}/\sqrt{2}$ with $X=12g^2+5\Omega_c^2$ and $Y=64g^4+24g^2\Omega_c^2+9\Omega_c^4$. Correspondingly, the eigenstates are given by
%
\begin{eqnarray*}
\Psi_{0+}^{(3)}&=&\frac{-\Omega_c^2}{\sqrt{6}g^2}|GG,3\rangle+\frac{\Omega_c}{g}|EG-,2\rangle+|EE,1\rangle\\
\Psi_{0-}^{(3)}&=&\frac{2g^2-\Omega_c^2}{\sqrt{6}g^2}|GG,3\rangle+\frac{\Omega_c}{g}|EG-,2\rangle+|MM,1\rangle\\
\Psi_{0}^{(3)}&=&\frac{\sqrt{2}g}{\Omega_c}|MG+,2\rangle+|EM-,1\rangle\\
\Psi_{1\pm}^{(3)}&=&\frac{\Omega_c}{\sqrt{2}g}|MG+,2\rangle\pm\frac{\sqrt{2g^2+\Omega_c^2}}{\sqrt{2}g}|EG+,2\rangle-|EM-,1\rangle\\
\Psi_{2\pm}^{(3)}&=&\pm\frac{Z+\sqrt{Y}}{2\sqrt{6}\Omega_c^2}|GG,3\rangle\pm\frac{\sqrt{X-\sqrt{Y}}(Z+\sqrt{Y})}{12\sqrt{2}g\Omega_c^2}|MG-,2\rangle\nonumber\\
& &+\frac{(Z-12g^2)+\sqrt{Y}}{12g\Omega_c}|EG-,2\rangle\pm\frac{\sqrt{X-\sqrt{Y}}}{2\Omega_c}|EM+,1\rangle\nonumber\\
& &+\frac{-Z+6\Omega_c^2-\sqrt{Y}}{6\Omega_c^2}|MM,1\rangle+|EE,1\rangle\\
\Psi_{3\pm}^{(3)}&=&\frac{Z-\sqrt{Y}}{2\sqrt{6}\Omega_c^2}|GG,3\rangle\pm\frac{\sqrt{X+\sqrt{Y}}(Z-\sqrt{Y})}{12\sqrt{2}g\Omega_c^2}|MG-,2\rangle\nonumber\\
& &+\frac{(Z-12g^2)-\sqrt{Y}}{12g\Omega_c}|EG-,2\rangle   \pm\frac{\sqrt{X+\sqrt{Y}}}{2\Omega_c}|EM+,1\rangle\nonumber\\
& &+\frac{-Z+6\Omega_c^2+\sqrt{Y}}{6\Omega_c^2}|MM,1\rangle+|EE,1\rangle
\end{eqnarray*}
%
with $Z=-8g^2+3\Omega_c^2$. 

The absorption of the third photon depends on the energy differences $\Delta E_{\rm 3ph}^{\pm}=\lambda^{(3)}_{3\pm}-3\lambda^{(2)}_{3\pm}/2$ at the detuning $\Delta_p=\lambda^{(2)}_{3\pm}/2$ and $\Delta E_{\rm 3ph}^{'\pm}=\lambda^{(3)}_{2\pm}-3\lambda^{(2)}_{2\pm}/2$ at the detuning $\Delta_p=\lambda^{(2)}_{2\pm}/2$, respectively. As shown in Fig.~\ref{Fig.S1}(b), the energy differences become larger as the control light increases. Thus, it is possible to prevent the system from absorbing the third photon by increasing the control field intensity, resulting a significant improvement of the three-photon blockade.
\bibliographystyle{apsrev4-1}
\bibliography{Refs}

\providecommand{\noopsort}[1]{}\providecommand{\singleletter}[1]{#1}%
\begin{thebibliography}{30}%
\makeatletter
\providecommand \@ifxundefined [1]{%
 \@ifx{#1\undefined}
}%
\providecommand \@ifnum [1]{%
 \ifnum #1\expandafter \@firstoftwo
 \else \expandafter \@secondoftwo
 \fi
}%
\providecommand \@ifx [1]{%
 \ifx #1\expandafter \@firstoftwo
 \else \expandafter \@secondoftwo
 \fi
}%
\providecommand \natexlab [1]{#1}%
\providecommand \enquote  [1]{``#1''}%
\providecommand \bibnamefont  [1]{#1}%
\providecommand \bibfnamefont [1]{#1}%
\providecommand \citenamefont [1]{#1}%
\providecommand \href@noop [0]{\@secondoftwo}%
\providecommand \href [0]{\begingroup \@sanitize@url \@href}%
\providecommand \@href[1]{\@@startlink{#1}\@@href}%
\providecommand \@@href[1]{\endgroup#1\@@endlink}%
\providecommand \@sanitize@url [0]{\catcode `\\12\catcode `\$12\catcode
  `\&12\catcode `\#12\catcode `\^12\catcode `\_12\catcode `\%12\relax}%
\providecommand \@@startlink[1]{}%
\providecommand \@@endlink[0]{}%
\providecommand \url  [0]{\begingroup\@sanitize@url \@url }%
\providecommand \@url [1]{\endgroup\@href {#1}{\urlprefix }}%
\providecommand \urlprefix  [0]{URL }%
\providecommand \Eprint [0]{\href }%
\providecommand \doibase [0]{http://dx.doi.org/}%
\providecommand \selectlanguage [0]{\@gobble}%
\providecommand \bibinfo  [0]{\@secondoftwo}%
\providecommand \bibfield  [0]{\@secondoftwo}%
\providecommand \translation [1]{[#1]}%
\providecommand \BibitemOpen [0]{}%
\providecommand \bibitemStop [0]{}%
\providecommand \bibitemNoStop [0]{.\EOS\space}%
\providecommand \EOS [0]{\spacefactor3000\relax}%
\providecommand \BibitemShut  [1]{\csname bibitem#1\endcsname}%
\let\auto@bib@innerbib\@empty
\bibitem [{\citenamefont {Imamo\ifmmode~\bar{g}\else \={g}\fi{}lu}\ \emph
  {et~al.}(1997)\citenamefont {Imamo\ifmmode~\bar{g}\else \={g}\fi{}lu},
  \citenamefont {Schmidt}, \citenamefont {Woods},\ and\ \citenamefont
  {Deutsch}}]{Imamoglu}%
  \BibitemOpen
  \bibfield  {author} {\bibinfo {author} {\bibfnamefont {A.}~\bibnamefont
  {Imamo\ifmmode~\bar{g}\else \={g}\fi{}lu}}, \bibinfo {author} {\bibfnamefont
  {H.}~\bibnamefont {Schmidt}}, \bibinfo {author} {\bibfnamefont
  {G.}~\bibnamefont {Woods}}, \ and\ \bibinfo {author} {\bibfnamefont
  {M.}~\bibnamefont {Deutsch}},\ }\href@noop {} {\bibfield  {journal} {\bibinfo
   {journal} {Phys. Rev. Lett.}\ }\textbf {\bibinfo {volume} {79}},\ \bibinfo
  {pages} {1467} (\bibinfo {year} {1997})}\BibitemShut {NoStop}%
\bibitem [{\citenamefont {Birnbaum}\ \emph {et~al.}(2005)\citenamefont
  {Birnbaum}, \citenamefont {Boca}, \citenamefont {Miller}, \citenamefont
  {Boozer}, \citenamefont {Northup},\ and\ \citenamefont {Kimble}}]{Birnbaum}%
  \BibitemOpen
  \bibfield  {author} {\bibinfo {author} {\bibfnamefont {K.~M.}\ \bibnamefont
  {Birnbaum}}, \bibinfo {author} {\bibfnamefont {A.}~\bibnamefont {Boca}},
  \bibinfo {author} {\bibfnamefont {R.}~\bibnamefont {Miller}}, \bibinfo
  {author} {\bibfnamefont {A.~D.}\ \bibnamefont {Boozer}}, \bibinfo {author}
  {\bibfnamefont {T.~E.}\ \bibnamefont {Northup}}, \ and\ \bibinfo {author}
  {\bibfnamefont {H.~J.}\ \bibnamefont {Kimble}},\ }\href@noop {} {\bibfield
  {journal} {\bibinfo  {journal} {Nature}\ }\textbf {\bibinfo {volume} {436}},\
  \bibinfo {pages} {87} (\bibinfo {year} {2005})}\BibitemShut {NoStop}%
\bibitem [{\citenamefont {Hoffman}\ \emph {et~al.}(2011)\citenamefont
  {Hoffman}, \citenamefont {Srinivasan}, \citenamefont {Schmidt}, \citenamefont
  {Spietz}, \citenamefont {Aumentado}, \citenamefont {T\"ureci},\ and\
  \citenamefont {Houck}}]{Hoffman}%
  \BibitemOpen
  \bibfield  {author} {\bibinfo {author} {\bibfnamefont {A.~J.}\ \bibnamefont
  {Hoffman}}, \bibinfo {author} {\bibfnamefont {S.~J.}\ \bibnamefont
  {Srinivasan}}, \bibinfo {author} {\bibfnamefont {S.}~\bibnamefont {Schmidt}},
  \bibinfo {author} {\bibfnamefont {L.}~\bibnamefont {Spietz}}, \bibinfo
  {author} {\bibfnamefont {J.}~\bibnamefont {Aumentado}}, \bibinfo {author}
  {\bibfnamefont {H.~E.}\ \bibnamefont {T\"ureci}}, \ and\ \bibinfo {author}
  {\bibfnamefont {A.~A.}\ \bibnamefont {Houck}},\ }\href@noop {} {\bibfield
  {journal} {\bibinfo  {journal} {Phys. Rev. Lett.}\ }\textbf {\bibinfo
  {volume} {107}},\ \bibinfo {pages} {053602} (\bibinfo {year}
  {2011})}\BibitemShut {NoStop}%
\bibitem [{\citenamefont {Liu}\ \emph {et~al.}(2014)\citenamefont {Liu},
  \citenamefont {Xu}, \citenamefont {Miranowicz},\ and\ \citenamefont
  {Nori}}]{Liu}%
  \BibitemOpen
  \bibfield  {author} {\bibinfo {author} {\bibfnamefont {Y.-x.}\ \bibnamefont
  {Liu}}, \bibinfo {author} {\bibfnamefont {X.-W.}\ \bibnamefont {Xu}},
  \bibinfo {author} {\bibfnamefont {A.}~\bibnamefont {Miranowicz}}, \ and\
  \bibinfo {author} {\bibfnamefont {F.}~\bibnamefont {Nori}},\ }\href@noop {}
  {\bibfield  {journal} {\bibinfo  {journal} {Physical Review A}\ }\textbf
  {\bibinfo {volume} {89}},\ \bibinfo {pages} {043818} (\bibinfo {year}
  {2014})}\BibitemShut {NoStop}%
\bibitem [{\citenamefont {Wang}\ \emph {et~al.}(2016)\citenamefont {Wang},
  \citenamefont {Miranowicz}, \citenamefont {Li},\ and\ \citenamefont
  {Nori}}]{Wang}%
  \BibitemOpen
  \bibfield  {author} {\bibinfo {author} {\bibfnamefont {X.}~\bibnamefont
  {Wang}}, \bibinfo {author} {\bibfnamefont {A.}~\bibnamefont {Miranowicz}},
  \bibinfo {author} {\bibfnamefont {H.-R.}\ \bibnamefont {Li}}, \ and\ \bibinfo
  {author} {\bibfnamefont {F.}~\bibnamefont {Nori}},\ }\href@noop {} {\bibfield
   {journal} {\bibinfo  {journal} {Physical Review A}\ }\textbf {\bibinfo
  {volume} {93}},\ \bibinfo {pages} {063861} (\bibinfo {year}
  {2016})}\BibitemShut {NoStop}%
\bibitem [{\citenamefont {Felicetti}\ \emph {et~al.}(2018)\citenamefont
  {Felicetti}, \citenamefont {Rossatto}, \citenamefont {Rico}, \citenamefont
  {Solano},\ and\ \citenamefont {Forn-D{\'\i}az}}]{Felicetti}%
  \BibitemOpen
  \bibfield  {author} {\bibinfo {author} {\bibfnamefont {S.}~\bibnamefont
  {Felicetti}}, \bibinfo {author} {\bibfnamefont {D.}~\bibnamefont {Rossatto}},
  \bibinfo {author} {\bibfnamefont {E.}~\bibnamefont {Rico}}, \bibinfo {author}
  {\bibfnamefont {E.}~\bibnamefont {Solano}}, \ and\ \bibinfo {author}
  {\bibfnamefont {P.}~\bibnamefont {Forn-D{\'\i}az}},\ }\href@noop {}
  {\bibfield  {journal} {\bibinfo  {journal} {Physical Review A}\ }\textbf
  {\bibinfo {volume} {97}},\ \bibinfo {pages} {013851} (\bibinfo {year}
  {2018})}\BibitemShut {NoStop}%
\bibitem [{\citenamefont {Faraon}\ \emph {et~al.}(2008)\citenamefont {Faraon},
  \citenamefont {Fushman}, \citenamefont {Englund}, \citenamefont {Stoltz},
  \citenamefont {Petroff},\ and\ \citenamefont {Vu{\v{c}}kovi{\'c}}}]{Faraon}%
  \BibitemOpen
  \bibfield  {author} {\bibinfo {author} {\bibfnamefont {A.}~\bibnamefont
  {Faraon}}, \bibinfo {author} {\bibfnamefont {I.}~\bibnamefont {Fushman}},
  \bibinfo {author} {\bibfnamefont {D.}~\bibnamefont {Englund}}, \bibinfo
  {author} {\bibfnamefont {N.}~\bibnamefont {Stoltz}}, \bibinfo {author}
  {\bibfnamefont {P.}~\bibnamefont {Petroff}}, \ and\ \bibinfo {author}
  {\bibfnamefont {J.}~\bibnamefont {Vu{\v{c}}kovi{\'c}}},\ }\href@noop {}
  {\bibfield  {journal} {\bibinfo  {journal} {Nature Physics}\ }\textbf
  {\bibinfo {volume} {4}},\ \bibinfo {pages} {859} (\bibinfo {year}
  {2008})}\BibitemShut {NoStop}%
\bibitem [{\citenamefont {Reinhard}\ \emph {et~al.}(2012)\citenamefont
  {Reinhard}, \citenamefont {Volz}, \citenamefont {Winger}, \citenamefont
  {Badolato}, \citenamefont {Hennessy}, \citenamefont {Hu},\ and\ \citenamefont
  {Imamo{\u{g}}lu}}]{Reinhard}%
  \BibitemOpen
  \bibfield  {author} {\bibinfo {author} {\bibfnamefont {A.}~\bibnamefont
  {Reinhard}}, \bibinfo {author} {\bibfnamefont {T.}~\bibnamefont {Volz}},
  \bibinfo {author} {\bibfnamefont {M.}~\bibnamefont {Winger}}, \bibinfo
  {author} {\bibfnamefont {A.}~\bibnamefont {Badolato}}, \bibinfo {author}
  {\bibfnamefont {K.~J.}\ \bibnamefont {Hennessy}}, \bibinfo {author}
  {\bibfnamefont {E.~L.}\ \bibnamefont {Hu}}, \ and\ \bibinfo {author}
  {\bibfnamefont {A.}~\bibnamefont {Imamo{\u{g}}lu}},\ }\href@noop {}
  {\bibfield  {journal} {\bibinfo  {journal} {Nature Photonics}\ }\textbf
  {\bibinfo {volume} {6}},\ \bibinfo {pages} {93} (\bibinfo {year}
  {2012})}\BibitemShut {NoStop}%
\bibitem [{\citenamefont {Rabl}(2011)}]{Rabi}%
  \BibitemOpen
  \bibfield  {author} {\bibinfo {author} {\bibfnamefont {P.}~\bibnamefont
  {Rabl}},\ }\href@noop {} {\bibfield  {journal} {\bibinfo  {journal} {Phys.
  Rev. Lett.}\ }\textbf {\bibinfo {volume} {107}},\ \bibinfo {pages} {063601}
  (\bibinfo {year} {2011})}\BibitemShut {NoStop}%
\bibitem [{\citenamefont {Ludwig}\ \emph {et~al.}(2012)\citenamefont {Ludwig},
  \citenamefont {Safavi-Naeini}, \citenamefont {Painter},\ and\ \citenamefont
  {Marquardt}}]{Ludwig}%
  \BibitemOpen
  \bibfield  {author} {\bibinfo {author} {\bibfnamefont {M.}~\bibnamefont
  {Ludwig}}, \bibinfo {author} {\bibfnamefont {A.~H.}\ \bibnamefont
  {Safavi-Naeini}}, \bibinfo {author} {\bibfnamefont {O.}~\bibnamefont
  {Painter}}, \ and\ \bibinfo {author} {\bibfnamefont {F.}~\bibnamefont
  {Marquardt}},\ }\href@noop {} {\bibfield  {journal} {\bibinfo  {journal}
  {Phys. Rev. Lett.}\ }\textbf {\bibinfo {volume} {109}},\ \bibinfo {pages}
  {063601} (\bibinfo {year} {2012})}\BibitemShut {NoStop}%
\bibitem [{\citenamefont {Liao}\ and\ \citenamefont {Nori}(2013)}]{Nori}%
  \BibitemOpen
  \bibfield  {author} {\bibinfo {author} {\bibfnamefont {J.-Q.}\ \bibnamefont
  {Liao}}\ and\ \bibinfo {author} {\bibfnamefont {F.}~\bibnamefont {Nori}},\
  }\href@noop {} {\bibfield  {journal} {\bibinfo  {journal} {Phys. Rev. A}\
  }\textbf {\bibinfo {volume} {88}},\ \bibinfo {pages} {023853} (\bibinfo
  {year} {2013})}\BibitemShut {NoStop}%
\bibitem [{\citenamefont {Hu}\ \emph {et~al.}(2015)\citenamefont {Hu},
  \citenamefont {Huang}, \citenamefont {Liao}, \citenamefont {Tian},\ and\
  \citenamefont {Goan}}]{Hu}%
  \BibitemOpen
  \bibfield  {author} {\bibinfo {author} {\bibfnamefont {D.}~\bibnamefont
  {Hu}}, \bibinfo {author} {\bibfnamefont {S.-Y.}\ \bibnamefont {Huang}},
  \bibinfo {author} {\bibfnamefont {J.-Q.}\ \bibnamefont {Liao}}, \bibinfo
  {author} {\bibfnamefont {L.}~\bibnamefont {Tian}}, \ and\ \bibinfo {author}
  {\bibfnamefont {H.-S.}\ \bibnamefont {Goan}},\ }\href@noop {} {\bibfield
  {journal} {\bibinfo  {journal} {Phys. Rev. A}\ }\textbf {\bibinfo {volume}
  {91}},\ \bibinfo {pages} {013812} (\bibinfo {year} {2015})}\BibitemShut
  {NoStop}%
\bibitem [{\citenamefont {Xie}\ \emph {et~al.}(2017)\citenamefont {Xie},
  \citenamefont {Liao}, \citenamefont {Shang}, \citenamefont {Ye},\ and\
  \citenamefont {Lin}}]{Xie}%
  \BibitemOpen
  \bibfield  {author} {\bibinfo {author} {\bibfnamefont {H.}~\bibnamefont
  {Xie}}, \bibinfo {author} {\bibfnamefont {C.-G.}\ \bibnamefont {Liao}},
  \bibinfo {author} {\bibfnamefont {X.}~\bibnamefont {Shang}}, \bibinfo
  {author} {\bibfnamefont {M.-Y.}\ \bibnamefont {Ye}}, \ and\ \bibinfo {author}
  {\bibfnamefont {X.-M.}\ \bibnamefont {Lin}},\ }\href@noop {} {\bibfield
  {journal} {\bibinfo  {journal} {Phys. Rev. A}\ }\textbf {\bibinfo {volume}
  {96}},\ \bibinfo {pages} {013861} (\bibinfo {year} {2017})}\BibitemShut
  {NoStop}%
\bibitem [{\citenamefont {Zhu}\ \emph {et~al.}(2018)\citenamefont {Zhu},
  \citenamefont {L\"u}, \citenamefont {Wan}, \citenamefont {Yin}, \citenamefont
  {Bin},\ and\ \citenamefont {Wu}}]{Wu}%
  \BibitemOpen
  \bibfield  {author} {\bibinfo {author} {\bibfnamefont {G.-L.}\ \bibnamefont
  {Zhu}}, \bibinfo {author} {\bibfnamefont {X.-Y.}\ \bibnamefont {L\"u}},
  \bibinfo {author} {\bibfnamefont {L.-L.}\ \bibnamefont {Wan}}, \bibinfo
  {author} {\bibfnamefont {T.-S.}\ \bibnamefont {Yin}}, \bibinfo {author}
  {\bibfnamefont {Q.}~\bibnamefont {Bin}}, \ and\ \bibinfo {author}
  {\bibfnamefont {Y.}~\bibnamefont {Wu}},\ }\href@noop {} {\bibfield  {journal}
  {\bibinfo  {journal} {Phys. Rev. A}\ }\textbf {\bibinfo {volume} {97}},\
  \bibinfo {pages} {033830} (\bibinfo {year} {2018})}\BibitemShut {NoStop}%
\bibitem [{\citenamefont {Hamsen}\ \emph {et~al.}(2017)\citenamefont {Hamsen},
  \citenamefont {Tolazzi}, \citenamefont {Wilk},\ and\ \citenamefont
  {Rempe}}]{Hamsen}%
  \BibitemOpen
  \bibfield  {author} {\bibinfo {author} {\bibfnamefont {C.}~\bibnamefont
  {Hamsen}}, \bibinfo {author} {\bibfnamefont {K.~N.}\ \bibnamefont {Tolazzi}},
  \bibinfo {author} {\bibfnamefont {T.}~\bibnamefont {Wilk}}, \ and\ \bibinfo
  {author} {\bibfnamefont {G.}~\bibnamefont {Rempe}},\ }\href@noop {}
  {\bibfield  {journal} {\bibinfo  {journal} {Physical review letters}\
  }\textbf {\bibinfo {volume} {118}},\ \bibinfo {pages} {133604} (\bibinfo
  {year} {2017})}\BibitemShut {NoStop}%
\bibitem [{\citenamefont {Schuster}\ \emph {et~al.}(2008)\citenamefont
  {Schuster}, \citenamefont {Kubanek}, \citenamefont {Fuhrmanek}, \citenamefont
  {Puppe}, \citenamefont {Pinkse}, \citenamefont {Murr},\ and\ \citenamefont
  {Rempe}}]{Schuster}%
  \BibitemOpen
  \bibfield  {author} {\bibinfo {author} {\bibfnamefont {I.}~\bibnamefont
  {Schuster}}, \bibinfo {author} {\bibfnamefont {A.}~\bibnamefont {Kubanek}},
  \bibinfo {author} {\bibfnamefont {A.}~\bibnamefont {Fuhrmanek}}, \bibinfo
  {author} {\bibfnamefont {T.}~\bibnamefont {Puppe}}, \bibinfo {author}
  {\bibfnamefont {P.~W.}\ \bibnamefont {Pinkse}}, \bibinfo {author}
  {\bibfnamefont {K.}~\bibnamefont {Murr}}, \ and\ \bibinfo {author}
  {\bibfnamefont {G.}~\bibnamefont {Rempe}},\ }\href@noop {} {\bibfield
  {journal} {\bibinfo  {journal} {Nature Physics}\ }\textbf {\bibinfo {volume}
  {4}},\ \bibinfo {pages} {382} (\bibinfo {year} {2008})}\BibitemShut {NoStop}%
\bibitem [{\citenamefont {Kubanek}\ \emph {et~al.}(2008)\citenamefont
  {Kubanek}, \citenamefont {Ourjoumtsev}, \citenamefont {Schuster},
  \citenamefont {Koch}, \citenamefont {Pinkse}, \citenamefont {Murr},\ and\
  \citenamefont {Rempe}}]{Kubanek}%
  \BibitemOpen
  \bibfield  {author} {\bibinfo {author} {\bibfnamefont {A.}~\bibnamefont
  {Kubanek}}, \bibinfo {author} {\bibfnamefont {A.}~\bibnamefont
  {Ourjoumtsev}}, \bibinfo {author} {\bibfnamefont {I.}~\bibnamefont
  {Schuster}}, \bibinfo {author} {\bibfnamefont {M.}~\bibnamefont {Koch}},
  \bibinfo {author} {\bibfnamefont {P.~W.~H.}\ \bibnamefont {Pinkse}}, \bibinfo
  {author} {\bibfnamefont {K.}~\bibnamefont {Murr}}, \ and\ \bibinfo {author}
  {\bibfnamefont {G.}~\bibnamefont {Rempe}},\ }\href@noop {} {\bibfield
  {journal} {\bibinfo  {journal} {Phys. Rev. Lett.}\ }\textbf {\bibinfo
  {volume} {101}},\ \bibinfo {pages} {203602} (\bibinfo {year}
  {2008})}\BibitemShut {NoStop}%
\bibitem [{\citenamefont {Souza}\ \emph {et~al.}(2013)\citenamefont {Souza},
  \citenamefont {Figueroa}, \citenamefont {Chibani}, \citenamefont
  {Villas-Boas},\ and\ \citenamefont {Rempe}}]{Souza}%
  \BibitemOpen
  \bibfield  {author} {\bibinfo {author} {\bibfnamefont {J.}~\bibnamefont
  {Souza}}, \bibinfo {author} {\bibfnamefont {E.}~\bibnamefont {Figueroa}},
  \bibinfo {author} {\bibfnamefont {H.}~\bibnamefont {Chibani}}, \bibinfo
  {author} {\bibfnamefont {C.}~\bibnamefont {Villas-Boas}}, \ and\ \bibinfo
  {author} {\bibfnamefont {G.}~\bibnamefont {Rempe}},\ }\href@noop {}
  {\bibfield  {journal} {\bibinfo  {journal} {Physical review letters}\
  }\textbf {\bibinfo {volume} {111}},\ \bibinfo {pages} {113602} (\bibinfo
  {year} {2013})}\BibitemShut {NoStop}%
\bibitem [{\citenamefont {Zhu}\ \emph {et~al.}(2017)\citenamefont {Zhu},
  \citenamefont {Yang},\ and\ \citenamefont {Agarwal}}]{Zhu}%
  \BibitemOpen
  \bibfield  {author} {\bibinfo {author} {\bibfnamefont {C.~J.}\ \bibnamefont
  {Zhu}}, \bibinfo {author} {\bibfnamefont {Y.~P.}\ \bibnamefont {Yang}}, \
  and\ \bibinfo {author} {\bibfnamefont {G.~S.}\ \bibnamefont {Agarwal}},\
  }\href@noop {} {\bibfield  {journal} {\bibinfo  {journal} {Phys. Rev. A}\
  }\textbf {\bibinfo {volume} {95}},\ \bibinfo {pages} {063842} (\bibinfo
  {year} {2017})}\BibitemShut {NoStop}%
\bibitem [{\citenamefont {Werner}\ and\ \citenamefont
  {Imamo\ifmmode~\bar{g}\else \={g}\fi{}lu}(1999)}]{Werner}%
  \BibitemOpen
  \bibfield  {author} {\bibinfo {author} {\bibfnamefont {M.~J.}\ \bibnamefont
  {Werner}}\ and\ \bibinfo {author} {\bibfnamefont {A.}~\bibnamefont
  {Imamo\ifmmode~\bar{g}\else \={g}\fi{}lu}},\ }\href@noop {} {\bibfield
  {journal} {\bibinfo  {journal} {Phys. Rev. A}\ }\textbf {\bibinfo {volume}
  {61}},\ \bibinfo {pages} {011801} (\bibinfo {year} {1999})}\BibitemShut
  {NoStop}%
\bibitem [{\citenamefont {Rebi\ifmmode~\acute{c}\else \'{c}\fi{}}\ \emph
  {et~al.}(2002)\citenamefont {Rebi\ifmmode~\acute{c}\else \'{c}\fi{}},
  \citenamefont {Parkins},\ and\ \citenamefont {Tan}}]{Rebic}%
  \BibitemOpen
  \bibfield  {author} {\bibinfo {author} {\bibfnamefont {S.}~\bibnamefont
  {Rebi\ifmmode~\acute{c}\else \'{c}\fi{}}}, \bibinfo {author} {\bibfnamefont
  {A.~S.}\ \bibnamefont {Parkins}}, \ and\ \bibinfo {author} {\bibfnamefont
  {S.~M.}\ \bibnamefont {Tan}},\ }\href@noop {} {\bibfield  {journal} {\bibinfo
   {journal} {Phys. Rev. A}\ }\textbf {\bibinfo {volume} {65}},\ \bibinfo
  {pages} {063804} (\bibinfo {year} {2002})}\BibitemShut {NoStop}%
\bibitem [{\citenamefont {Liao}\ \emph {et~al.}(2010)\citenamefont {Liao},
  \citenamefont {Law} \emph {et~al.}}]{Liao}%
  \BibitemOpen
  \bibfield  {author} {\bibinfo {author} {\bibfnamefont {J.-Q.}\ \bibnamefont
  {Liao}}, \bibinfo {author} {\bibfnamefont {C.}~\bibnamefont {Law}},  \emph
  {et~al.},\ }\href@noop {} {\bibfield  {journal} {\bibinfo  {journal}
  {Physical Review A}\ }\textbf {\bibinfo {volume} {82}},\ \bibinfo {pages}
  {053836} (\bibinfo {year} {2010})}\BibitemShut {NoStop}%
\bibitem [{\citenamefont {Miranowicz}\ \emph {et~al.}(2013)\citenamefont
  {Miranowicz}, \citenamefont {Paprzycka}, \citenamefont {Liu}, \citenamefont
  {Bajer},\ and\ \citenamefont {Nori}}]{Miranowicz}%
  \BibitemOpen
  \bibfield  {author} {\bibinfo {author} {\bibfnamefont {A.}~\bibnamefont
  {Miranowicz}}, \bibinfo {author} {\bibfnamefont {M.}~\bibnamefont
  {Paprzycka}}, \bibinfo {author} {\bibfnamefont {Y.-x.}\ \bibnamefont {Liu}},
  \bibinfo {author} {\bibfnamefont {J.}~\bibnamefont {Bajer}}, \ and\ \bibinfo
  {author} {\bibfnamefont {F.}~\bibnamefont {Nori}},\ }\href@noop {} {\bibfield
   {journal} {\bibinfo  {journal} {Physical Review A}\ }\textbf {\bibinfo
  {volume} {87}},\ \bibinfo {pages} {023809} (\bibinfo {year}
  {2013})}\BibitemShut {NoStop}%
\bibitem [{\citenamefont {Liew}\ and\ \citenamefont {Savona}(2010)}]{Liew}%
  \BibitemOpen
  \bibfield  {author} {\bibinfo {author} {\bibfnamefont {T.~C.~H.}\
  \bibnamefont {Liew}}\ and\ \bibinfo {author} {\bibfnamefont {V.}~\bibnamefont
  {Savona}},\ }\href@noop {} {\bibfield  {journal} {\bibinfo  {journal} {Phys.
  Rev. Lett.}\ }\textbf {\bibinfo {volume} {104}},\ \bibinfo {pages} {183601}
  (\bibinfo {year} {2010})}\BibitemShut {NoStop}%
\bibitem [{\citenamefont {Majumdar}\ \emph {et~al.}(2012)\citenamefont
  {Majumdar}, \citenamefont {Bajcsy}, \citenamefont {Rundquist},\ and\
  \citenamefont {Vu\ifmmode \check{c}\else
  \v{c}\fi{}kovi\ifmmode~\acute{c}\else \'{c}\fi{}}}]{Majumdar}%
  \BibitemOpen
  \bibfield  {author} {\bibinfo {author} {\bibfnamefont {A.}~\bibnamefont
  {Majumdar}}, \bibinfo {author} {\bibfnamefont {M.}~\bibnamefont {Bajcsy}},
  \bibinfo {author} {\bibfnamefont {A.}~\bibnamefont {Rundquist}}, \ and\
  \bibinfo {author} {\bibfnamefont {J.}~\bibnamefont {Vu\ifmmode \check{c}\else
  \v{c}\fi{}kovi\ifmmode~\acute{c}\else \'{c}\fi{}}},\ }\href@noop {}
  {\bibfield  {journal} {\bibinfo  {journal} {Phys. Rev. Lett.}\ }\textbf
  {\bibinfo {volume} {108}},\ \bibinfo {pages} {183601} (\bibinfo {year}
  {2012})}\BibitemShut {NoStop}%
\bibitem [{\citenamefont {Bamba}\ \emph {et~al.}(2011)\citenamefont {Bamba},
  \citenamefont {Imamo\ifmmode~\breve{g}\else \u{g}\fi{}lu}, \citenamefont
  {Carusotto},\ and\ \citenamefont {Ciuti}}]{Bamba}%
  \BibitemOpen
  \bibfield  {author} {\bibinfo {author} {\bibfnamefont {M.}~\bibnamefont
  {Bamba}}, \bibinfo {author} {\bibfnamefont {A.}~\bibnamefont
  {Imamo\ifmmode~\breve{g}\else \u{g}\fi{}lu}}, \bibinfo {author}
  {\bibfnamefont {I.}~\bibnamefont {Carusotto}}, \ and\ \bibinfo {author}
  {\bibfnamefont {C.}~\bibnamefont {Ciuti}},\ }\href@noop {} {\bibfield
  {journal} {\bibinfo  {journal} {Phys. Rev. A}\ }\textbf {\bibinfo {volume}
  {83}},\ \bibinfo {pages} {021802} (\bibinfo {year} {2011})}\BibitemShut
  {NoStop}%
\bibitem [{\citenamefont {Gerace}\ and\ \citenamefont {Savona}(2014)}]{Gerace}%
  \BibitemOpen
  \bibfield  {author} {\bibinfo {author} {\bibfnamefont {D.}~\bibnamefont
  {Gerace}}\ and\ \bibinfo {author} {\bibfnamefont {V.}~\bibnamefont
  {Savona}},\ }\href@noop {} {\bibfield  {journal} {\bibinfo  {journal} {Phys.
  Rev. A}\ }\textbf {\bibinfo {volume} {89}},\ \bibinfo {pages} {031803}
  (\bibinfo {year} {2014})}\BibitemShut {NoStop}%
\bibitem [{\citenamefont {Flayac}\ and\ \citenamefont {Savona}(2017)}]{Flayac}%
  \BibitemOpen
  \bibfield  {author} {\bibinfo {author} {\bibfnamefont {H.}~\bibnamefont
  {Flayac}}\ and\ \bibinfo {author} {\bibfnamefont {V.}~\bibnamefont
  {Savona}},\ }\href@noop {} {\bibfield  {journal} {\bibinfo  {journal} {Phys.
  Rev. A}\ }\textbf {\bibinfo {volume} {96}},\ \bibinfo {pages} {053810}
  (\bibinfo {year} {2017})}\BibitemShut {NoStop}%
\bibitem [{\citenamefont {Vaneph}\ \emph {et~al.}(2018)\citenamefont {Vaneph},
  \citenamefont {Morvan}, \citenamefont {Aiello}, \citenamefont {F\'echant},
  \citenamefont {Aprili}, \citenamefont {Gabelli},\ and\ \citenamefont
  {Est\`eve}}]{Vaneph}%
  \BibitemOpen
  \bibfield  {author} {\bibinfo {author} {\bibfnamefont {C.}~\bibnamefont
  {Vaneph}}, \bibinfo {author} {\bibfnamefont {A.}~\bibnamefont {Morvan}},
  \bibinfo {author} {\bibfnamefont {G.}~\bibnamefont {Aiello}}, \bibinfo
  {author} {\bibfnamefont {M.}~\bibnamefont {F\'echant}}, \bibinfo {author}
  {\bibfnamefont {M.}~\bibnamefont {Aprili}}, \bibinfo {author} {\bibfnamefont
  {J.}~\bibnamefont {Gabelli}}, \ and\ \bibinfo {author} {\bibfnamefont
  {J.}~\bibnamefont {Est\`eve}},\ }\href@noop {} {\bibfield  {journal}
  {\bibinfo  {journal} {Phys. Rev. Lett.}\ }\textbf {\bibinfo {volume} {121}},\
  \bibinfo {pages} {043602} (\bibinfo {year} {2018})}\BibitemShut {NoStop}%
\bibitem [{\citenamefont {Snijders}\ \emph {et~al.}(2018)\citenamefont
  {Snijders}, \citenamefont {Frey}, \citenamefont {Norman}, \citenamefont
  {Flayac}, \citenamefont {Savona}, \citenamefont {Gossard}, \citenamefont
  {Bowers}, \citenamefont {van Exter}, \citenamefont {Bouwmeester},\ and\
  \citenamefont {L\"offler}}]{Snijders}%
  \BibitemOpen
  \bibfield  {author} {\bibinfo {author} {\bibfnamefont {H.~J.}\ \bibnamefont
  {Snijders}}, \bibinfo {author} {\bibfnamefont {J.~A.}\ \bibnamefont {Frey}},
  \bibinfo {author} {\bibfnamefont {J.}~\bibnamefont {Norman}}, \bibinfo
  {author} {\bibfnamefont {H.}~\bibnamefont {Flayac}}, \bibinfo {author}
  {\bibfnamefont {V.}~\bibnamefont {Savona}}, \bibinfo {author} {\bibfnamefont
  {A.~C.}\ \bibnamefont {Gossard}}, \bibinfo {author} {\bibfnamefont {J.~E.}\
  \bibnamefont {Bowers}}, \bibinfo {author} {\bibfnamefont {M.~P.}\
  \bibnamefont {van Exter}}, \bibinfo {author} {\bibfnamefont {D.}~\bibnamefont
  {Bouwmeester}}, \ and\ \bibinfo {author} {\bibfnamefont {W.}~\bibnamefont
  {L\"offler}},\ }\href@noop {} {\bibfield  {journal} {\bibinfo  {journal}
  {Phys. Rev. Lett.}\ }\textbf {\bibinfo {volume} {121}},\ \bibinfo {pages}
  {043601} (\bibinfo {year} {2018})}\BibitemShut {NoStop}%
\end{thebibliography}%

\end{document}